\documentclass[journal,twocolumn]{IEEEtran}
\usepackage{amssymb}
\usepackage{amsfonts}
\usepackage{amsmath}
\usepackage{algorithm}
\usepackage{algorithmic}
\usepackage{subeqnarray}
\usepackage{cases}
\usepackage{mathrsfs}
\usepackage{algorithm}
\usepackage{algorithmic}
\usepackage{balance}

\usepackage{graphicx,subfigure,amsmath,amssymb,cite}
\newcommand{\tabincell}[2]{\begin{tabular}{@{}#1@{}}#2\end{tabular}}

\usepackage[dvips]{color}
\usepackage{float}
\ifCLASSINFOpdf
\else
\fi
\begin{document}

\title{\huge Regional Robust Secure Precise Wireless Transmission Design for Multi-user UAV Broadcasting System
}

\author{Tong Shen,~Tingting Liu, \emph{Member, IEEE},~Yan Lin, \emph{Member, IEEE},~Yongpeng Wu, \emph{Senior Member, IEEE},\\
~Feng Shu, \emph{Member, IEEE},~and Zhu Han, \emph{Fellow, IEEE}

\thanks{T. Shen,~T. Liu,~Y. Lin, and F. Shu are with the School of Electronic and Optical Engineering, Nanjing University of Science and Technology, Nanjing 210094, China (e-mail: shentong0107@163.com, liutt@njit.edu.cn, yanlin@njust.edu.cn, shufeng0101@163.com). }
\thanks{Y. Wu is with the Department of Electronic Engineering, Shanghai Jiao Tong University, Shanghai 200240, China (E-mail: yongpeng.wu@sjtu.edu.cn).}
\thanks{Z. Han is with the University of Houston, Houston, TX 77004 USA, and also with the Department of Computer Science and Engineering, Kyung Hee University, Seoul 02447, South Korea (e-mail: zhan2@uh.edu).
}
}

\maketitle

\begin{abstract}
In this paper, two regional robust secure precise wireless transmission (SPWT) schemes for multi-user unmanned aerial vehicle (UAV) :1) regional signal-to-leakage-and-noise ratio (SLNR) and artificial-noise-to-leakage-and-noise ratio (ANLNR) (R-SLNR-ANLNR) maximization and 2) point SLNR and ANLNR (P-SLNR-ANLNR) maximization, are proposed to tackle with the estimation errors of the target users' location. In SPWT system, the estimation error for SPWT can not be ignored. However the conventional robust methods in secure wireless communications optimize the beamforming vector in the desired positions only in statistical means and can not guarantee the security for each symbol. Proposed regional robust schemes are designed for optimizing the secrecy performance in the whole error region around the estimated location. Specifically, with known maximal estimation error, we define target region and wiretap region. Then design an optimal beamforming vector and an artificial noise projection matrix, which achieve the confidential signal in the target area having the maximal power while only few signal power is conserved in the potential wiretap region. Instead of considering the statistical distributions of the estimated errors into optimization, we optimize the SLNR and ANLNR of the whole target area, which significantly decreases the complexity. Moreover, the proposed schemes can ensure that the desired users are located in the optimized region, which are more practical than conventional methods. Simulation results show that our proposed regional robust SPWT design is capable of substantially improving the secrecy rate compared to the conventional non-robust method. The P-SLNR-ANLNR maximization-based method has the comparable secrecy performance with a lower complexity than that of the R-SLNR-ANLNR maximization-based method.
\end{abstract}
\section{Introduction}
In the past decade, physical layer security (PLS) in wireless communications \cite{Zou2016Relay,Wang2012Distributed,Wu2017Secure,ChenX2017,Zhu2016Security,zhou2019,ZhouTSP,JZhao2019,Ni2019,JZhao2018} have explosively developed because it is an alternative to encryption in the higher layer. In \cite{Goel2008Guaranteeing}, \textit{Goel et al.} first proposed artificial noise (AN) to improve the security performance in PLS. In \cite{Yan2016Artificial,Zhao2016Anti,Zhao2016Physical}, AN-aided secure transmissions were proposed for enhancing the security performance in wireless transmission. In \cite{Han1}, a secure wireless transmission with multiple assisting jammers by maximizing secrecy rate and optimizing power allocation was proposed. A physical layer approach for secure green communication was proposed by untrusted two-way relaying in \cite{Han2,Han3}. As another promising PLS technique, direction modulation (DM) \cite{Daly2009Directional,Yuan2014A,Yuan2015A,Yuan2014B}, has attracted extensive attention since it is capable of projecting useful signals only into a predetermined direction, while making the constellation of the signal in other directions distort. For instance, \textit{Daly et al.} of \cite{Daly2009Directional} presented a DM technique for modulation based upon a phased array. As a further extension, Yuan \textit{et al.} proposed an orthogonal vector approach in \cite{Yuan2014A,Yuan2015A,Yuan2014B}, which allows the imposed artificial orthogonal noise to bear on the DM analysis and synthesis. Additionally, by taking estimation error into consideration, some robust DM algorithms are proposed in various distributions of estimation error for PLS. For instance, in \cite{Hu2016Robust}, \textit{Hu et al.} considered the direction angle estimation error in uniform distribution and proposed a robust algorithm in a single-user DM scenario. Then, \textit{Shu et al.} in \cite{Shu2017Robust} proposed a multi-user robust scheme in the DM system with the direction angle estimation error with a Gaussian distribution, while \textit{Gui et al.} of \cite{gui2018robust} proposed a robust DM method with Von Mises distributed direction angle estimation error in multi-cast scenario. Another interesting proposal is a novel robust DM scheme of \cite{Shu2017Robusts} based on main-lobe-integration maximization. However, the security of these DM techniques is guaranteed only in the direction dimension, while the eavesdroppers in the desired direction may also receive the confidential messages from the channel, even if in different distances, which may bring a serious challenging security problem.

To address the aforementioned problem, the authors of \cite{Liu2016The,Hu2017SPWT,Zhu2017Secure,Shu2018SPWT} proposed DM based secure precise wireless transmission (SPWT) techniques which transmit confidential messages to the users with exact direction and distance. To elaborate, \textit{Liu et al.} in \cite{Liu2016The} proposed a novel DM scheme with random frequency diverse array (RFDA) which can produce a joint direction and distance controllable beam-pattern. In \cite{Hu2017SPWT}, \textit{Hu et al.} proposed a RFDA scheme combined with AN, which can impose the interference to the undesired receivers. Moreover, \textit{zhu et al.} in \cite{Zhu2017Secure} also proposed another SPWT scheme with the aid of cooperative relays.
As a further advance, in \cite{Shu2018SPWT}, \textit{Shu et al.} proposed a SPWT scheme by using random subcarrier selection and orthogonal frequency division multiplexing (OFDM), which has a low system complexity and reduce budget.
Nevertheless, all the aforementioned SPWT schemes above assume precise direction angle and distance or precise channel state information (CSI) between the transmitter and desired receivers. However, in practical scenarios, the precise position information or CSI is intractable to obtain, no matter which estimation method is utilized, including the most widely used direction estimation algorithm, MUSIC and Capon, or the distance estimation algorithm of received signal strength indication (RSSI), all of which still exist estimation error. To the best of our knowledge, research on the practical SPWT which considers the estimation error are still in its infancy. Although, there are some robust algorithms proposed in the DM system, most are not practical, since they assume that the estimation error distributions are known. And in vehicular networks, the communication requires high accuracy for each transmitted symbol, a statistical average optimal beamforming can not guarantee SPWT for each transmitted symbol.

Against the above backdrop, in this paper, for the first time, we consider a robust SPWT design for a multi-user unmanned aerial vehicle (UAV) broadcasting scenario and propose two regional robust multi-users SPWT schemes. Explicitly, both methods consider the direction angle and distance estimation error relying on the knowledge of their maximal estimation error range of direction angle and distance, rather than on the knowledge of their distributions. Then, by maximizing the confidential message energy in the estimation error region, the receivers, who are still in the estimation error region, can obtain the confidential message, regardless of the existing estimation errors. The proposed methods are more practical compared to the algorithm with estimation error in a statistical distribution. Our main contributions are summarized as follows:

\begin{enumerate}
\item A regional signal-to-leakage-noise ratio and artificial-noise-to-leakage-and-noise ratio (R-SLNR-ANLNR) maximization-based robust SPWT scheme is proposed with the aim of maximizing the region SLNR and ANLNR . And this is the first time that a robust scheme is proposed in UAV SPWT system. As the desired users must be located in the estimation error region, they can receive the more energy of the confidential message. Otherwise, they may receive rare signal energy even though with the aid of beamforming and AN. In this scheme, a maximizing the minimal SLNR problem is considered for guaranteeing the security performance of each user.
\item Considering the facts of a large amount of calculations and high complexity in the above scheme, we propose a low-complexity point SLNR and ANLNR (P-SLNR-ANLNR) maximization-based scheme by maximizing the performance of the sampled points. We first assume the main-lobe range is larger than the estimation error range, due to the fact that the lobe width is dependent on the antenna number and on the sub-carrier bandwidth. Then, we choose to sample several points in the estimation error region of each user. Through calculating the point energy sum rather than the whole region energy sum, the proposed P-SLNR-ANLNR maximization-based scheme can significantly decrease the complexity.
\item By numerical simulations, our proposed schemes are capable of achieving the transmitted confidential message energy mapped only in the specific estimation error regions of the desired users. Moreover, we observe that the secrecy performance of the P-SLNR-ANLNR maximization-based scheme is related to the estimation error and the main-lobe width.
\end{enumerate}

The rest of the paper is organized as follows: In Section II, the system model is introduced. In Section III, a regional robust SPWT scheme based on maximizing the SLNR of the desired user region is proposed. In Section IV, a low complexity regional robust SPWT method based on P-SLNR maximization is presented. Simulation results are shown in Section V. Finally, our conclusions are drawn in Section VI.

Notations: In this paper, we utilize letters of bold upper case, bold low case and low case to denote matrices, vectors, and scalars, respectively. Given a complex vector or matrix, $(\cdot)^T$, $(\cdot)^H$ and $tr(\cdot)$ denote the transpose, conjugate transpose, and trace, respectively.$\mathbf{I}_N $ refers to the $N\times N$ identity matrix. For a given set, $\overline{(\cdot)}$ denote the complement set.
\section{system model}
\begin{figure}[t]
\centering
\includegraphics[width=0.50\textwidth]{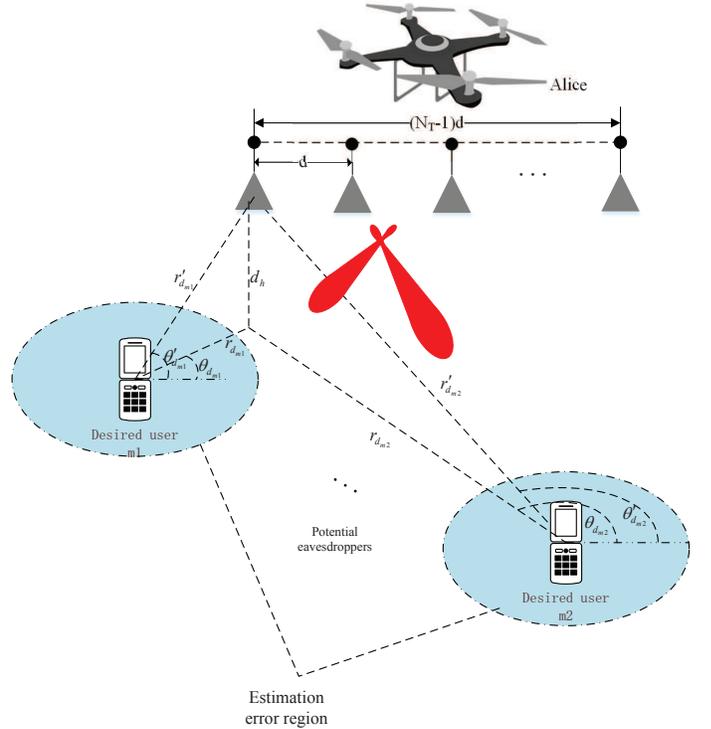}\\
\caption{Schematic diagram of the proposed scheme.}\label{system.eps}
\end{figure}

As shown in Fig.\ref{system.eps}, we consider a multi-user broadcasting system. A transmitter Alice is an UAW equipped with a linear $N_T$ transmit antenna array, and there are $M$ desired vehicular users equipped with single antenna, while the number and the location of eavesdroppers are unknown for Alice. We assume that the channel is the line-of-sight (LoS) channel, and the receivers are far from the transmit antenna array. Furthermore, eavesdroppers only exist out of the receivers' secure regions, which can be justified when eavesdroppers are convert and they need to keep enough distance from users to prevent detection. 
Thus, the signal transmitted by antenna  $n$ can be represented by
\begin{align}\label{s_n}
{s_n} &= \sqrt {{\alpha_1 {P}}} xe{}^{j {\phi _n} } + \sqrt {{\alpha_2 {P}}}  {w_n},
\end{align}
where $x$ is the modulated transmitting data symbol with $\mathbb{E}\left\{ {{x^*}x} \right\} = 1$, $w_n$ is the AN mapped to the $n$-th antenna, $P$ is the total transmit power of all antennas, $\alpha_1$ and $\alpha_2$ are the power allocation (PA) factors which satisfy the identity equation constraint $\alpha _1+\alpha _2=1$, and $\phi _n$ is the initial phase which we have to design. In this paper, we assume $\alpha_1$ and $\alpha_2$ are predesigned \cite{Wan2018}.
We choose the first antenna of the array on UAV as the reference antenna. All desired users and eavesdroppers are assumed to be on the ground and have the same height while the height of UAV is $d_h$. 

Establish the circular coordinate system with the projection of the first antenna on the ground as the origin, the location of an arbitrarily receiver is given by $(\theta,r)$. Then, the distance from the receiver to the first antenna of UAV can be expressed as $r^{'}=\sqrt{{r}^2+d^2_h}$. The direction between the ray from the user to the first antenna of UAV and the antenna array is expressed as
\begin{align}\label{theta}
 \theta^{'}=\arccos(\frac{r}{r^{'}}\cos \theta ),
\end{align}
since 
\begin{align}\label{2}
r^{'} \cos \theta^{'} = r \cos \theta.
\end{align}
Then the distance $r_{n}$ from the $n$-th antenna to the receiver can be expressed as
\begin{align}\label{R_n}
{r^{'}_{n}} = r^{'} - \left( {n - 1} \right)d\cos \theta^{'} ,
\end{align}
where $d$ denotes the elements spacing. As the reference antenna is the first element of the antenna array, we have $r_{1}=r$. The reference phase at the first antenna to be received is given by
\begin{align}
{\varphi _0}\left( {\theta ,r} \right) =2\pi{f_c}\frac{{{r^{'}}}}{c}\ ,
\end{align}
where $f_c$ is the central carrier frequency.
Similarly, the phase of the transmit signal at the $n$-th antenna received by the $m$-th user can be expressed as
\begin{align}
{\varphi _n}\left( {\theta ,r} \right) =2\pi({f_c+k_n\Delta f})\frac{{{r^{'}_{n}}}}{c}\ .
\end{align}
where ${f_c} + {k_n}\Delta f$ is the subcarrier mapped into the $n$-th antenna, and selected from a subcarriers set
\begin{align}
S_{sub}=\{f_i|f_i=f_c+i\Delta f,(i=0,1,\ldots,N-1)\}.
\end{align}
Herein, $N$ is the number of the total subcarriers, and the total bandwidth is $B=N\Delta f$.
Then the phase shifting corresponding to the reference phase of the $n$-th element is expressed as
\begin{align}\label{psi}
{\psi _n}\left( {\theta ,r} \right) &= {\rm{ }}{{ 2\pi \left( {{\rm{ }}{f_c} + {k_n}\Delta f} \right){\rm{ }}{{r^{'}_{n}}}/{c}}}-{\varphi _0}\left( {\theta ,r} \right),
\end{align}
Note that the subcarriers are randomly selected with randomization processing \cite{Shen2019}.
Then, the received signal at $\left( {\theta ,r} \right)$ can be represented as
 \begin{align}\label{Y}
y\left( {\theta ,r} \right)=\sqrt {\alpha_1 {P}}{\mathbf{h}^H}\mathbf{v}x + \sqrt {\alpha_2 {P}}{\mathbf{h}^H}\mathbf{w} + n ,
\end{align}
where $\mathbf{h}$ is the normalized steering vector given by
\begin{align}\label{h}
\mathbf{h} = \frac{1}{{{\sqrt{N_T}}}} {\left[ {{e^{j{\psi _1}\left( {\theta ,r} \right)}},{e^{j{\psi _2}\left( {\theta ,r} \right)}}, \cdots ,{e^{j{\psi _{{N_T}}}\left( {\theta ,r} \right)}}} \right]^T}.
\end{align}
Furthermore, $\mathbf{v}$, where $\mathbf{v}^H\mathbf{v}$=1, is the beamforming vector designed to achieve phase alignment,
and $n$ is the additive white Gaussian noise (AWGN) distributed as $n\sim\mathcal{CN}(0,\sigma^2)$.
Accordingly, the received signals along the $m$-th desired user can be expressed as
\begin{align}
y\left( {{\theta _{{d_m}}},{r_{{d_m}}}} \right) = \sqrt {{\alpha _1}{P}} {\bf{h}}_{{d_m}}^H{\bf{v}}x + \sqrt {{\alpha _2}{P}} {\bf{h}}_{{d_m}}^H{\bf{w}} + {n_{{d_m}}},
\end{align}
where $d_m$ denotes the $m$-th desired user, and $\mathbf{h_{d_m}}$ is obtained by replacing $\left( {\theta ,r} \right)$ with $\left( {\theta_{d_m} ,r_{d_m}} \right)$ in (\ref{psi}), $n_{{d_m}}$ is the AWGN distributed as $n_{{d_m}}\sim\mathcal{CN}(0,\sigma_{d_m}^2)$. Additionally, the received signals along the undesired position where eavesdropper may be located can be expressed as
\begin{align}
y\left( {{\theta _{{e_k}}},{r_{{e_k}}}} \right) = \sqrt {{\alpha _1}{P}} {\bf{h}}_{{e_k}}^H{\bf{v}}x + \sqrt {{\alpha _2}{P}} {\bf{h}}_{{e_k}}^H{\bf{w}} + {n_{{e_k}}},
\end{align}
where $e_k$ denotes the $k$-th eavesdropper, and $\mathbf{h_{e_k}}$ is obtained by replacing $\left( {\theta ,r} \right)$ with $\left( {\theta_{e_k} ,r_{e_k}} \right)$ in (\ref{h}), and $n_{{e_k}}\sim\mathcal{CN}(0,\sigma_{e_k}^2)$. For simplicity, we assume $\sigma_{d_m}^2=\sigma_{e_k}^2=\sigma^2$. Then, our aim is to find the optimal $\mathbf{v}$ and $\mathbf{w}$ for achieving the best secrecy performance. However, in general the transmitter does not have the perfect knowledge of the desired receivers and neither have that of the eavesdroppers. All the knowledge of direction and distance is measured by diverse estimation methods and hence results in estimation errors. Let us define the estimated position of the $m$-th user $(\theta_{d_m},r_{d_m})$ as $(\hat{\theta}_{d_m},\hat{r}_{d_m})$, but our designed $\mathbf{v}$ and $\mathbf{w}$ with the estimated knowledge which may bring performance loss. Thus, we describe a robust scenario in the following.

Consider that the positions of the legitimate users can be obtained by some well-known estimation methods. For instance, the direction angle can be estimated by some classical direction estimation methods such as multiple signal classification (MUSIC) \cite{Schmidt,Qin2018}, and the distance, from the transmitter to the receivers, may be estimated by some typical methods such as channel state information (CSI) based localization method \cite{gui2018cramer}, which is available in both indoor and outdoor scenarios with high-distance-resolution. Practically the estimation error and its corresponding error distribution have to be considered in the system, which may increase the complexity of the SPWT algorithm. In this paper, we propose the regional robust methods to simplify the algorithm with only the maximal estimation errors, this scheme maps the useful information into a region around the estimated position.

First, we assume that the maximal estimation errors of angle and distance are defined as $\Delta\theta_{\mathrm{max}}$ and $\Delta r_{\mathrm{max}}$. The maximal estimation errors are depend on the measurement method we chooses.  In this paper, we assume that with these measurement schemes, the maximal estimation errors are known. Then we define the $m$-th desired region given by
\begin{align}
\mathrm{Area}_{d_m}=\{(\theta,r)|\hat{\theta}_{d_m}-\Delta\theta_{\mathrm{max}}\leq\theta\leq\hat{\theta}_{d_m}+\Delta\theta_{\mathrm{max}},\nonumber\\
\hat{r}_{d_m}-\Delta r_{\mathrm{max}} \leq R \leq \hat{r}_{d_m}+\Delta r_{\mathrm{max}}\}.
\end{align}
Generally, the position of eavesdroppers are usually unknown, which means a potential eavesdropper may exist in everywhere out of the desired regions. Thus we can define the area out of the desired regions as the wiretap region, given by
\begin{align}
\mathrm{Area}_{e}=\overline{\bigcup\limits_{m = 1}^M \mathrm{Area}_{d_m}}.
\end{align}

In the following sections, we will design a feasible beamforming vector $\mathbf{v}$ and an artificial noise vector $\mathbf{w}$ to make the transmitted confidential signal energy mainly concentrate on the desired region as a regional robust SWPT.
\section{Method based on regional SLNR maximization}

This section is to propose a regional robust method based on optimizing the product of signal-to-leakeage-and-noise ratio (SLNR) and artificial-noise-to-leakage-and-noise ratio (ANLNR). The SLNR refers to the ratio between the confidential message energy at desired user and  the confidential message energy at eavesdroppers. The ANLNR refers to the ratio between the artificial noise energy at eavesdroppers and  the artificial noise energy at desired users. A higher SLNR means the more confidential message energy concentrate on the desired user and a higher ANLNR means artificial noise more concentrate on eavesdroppers and less effect to desired users.
The SLNR for the $m$-th user is defined as
\begin{align}\label{SLNR}
\mathrm{SLNR}_m = \frac{{{{\alpha _1}P{\bf{h}}_{{d_m}}^H{\bf{v}}{{\bf{v}}^H}{{\bf{h}}_{{d_m}}}} }}{{\sum\limits_{k = 1}^K {\left( {{\alpha _1}P{\bf{h}}_{{e_k}}^H{\bf{v}}{{\bf{v}}^H}{{\bf{h}}_{{e_k}}} } \right)+ \sigma^2} }},
\end{align}
where $K$ is the number of eavesdroppers. And the ANLNR is defined as
\begin{align}\label{ANLNR}
\mathrm{ANLNR_m}= \frac{\sum\limits_{k = 1}^K{{{\alpha _2}P{\bf{h}}_{{e_k}}^H{\bf{w}}{{\bf{w}}^H}{{\bf{h}}_{{e_k}}}} }}{{ { {{\alpha _2}P{\bf{h}}_{{d_m}}^H{\bf{w}}{{\bf{w}}^H}{{\bf{h}}_{{d_m}}} } + \sigma^2} }},
\end{align}
For single user and eavesdropper, in the medium and high SNR region, $\log(\mathrm{SLNR}\cdot\mathrm{{ANLNR}})\approx \log(\mathrm{SINR_B})-\log(\mathrm{SINR_E})\approx \mathrm{SR}$, where $\mathrm{SR}$ is the secrecy rate. In other words, maximizing the product is equivalent to maximizing SR. In order to improve the security performance, both the SLNR at Bob and the ANLNR at Eve need to be maximized. First, we optimize the beamforming vector $v$ by maximizing the SLNR of the SLNR-minimum desired user.

Taking the estimation errors into consideration, we define the $m$-th user's regional SLNR as
\begin{align}
\hat{\mathrm{SLNR}}_m=\frac{{{\alpha _1}P\text{tr}\left\{ {{{\mathbf{v}}^H}\hat{\mathbf{R}}_{d_m}{{\mathbf{v}}}} \right\}}}{{\text{tr}\left\{ {{{\mathbf{v}}^H}\left( {{\alpha _1}P\hat{\mathbf{R}}_e + \sigma^2{\mathbf{I}_N}} \right){{\mathbf{v}}}} \right\}}},
\end{align}
where
\begin{align}\label{R_d_m}
\hat{\mathbf{R}}_{d_m}=\mathop{{\int\!\!\!\!\!\int}\mkern-21mu \bigcirc}\limits_{Area_{d_m}}
 {\mathbf{h}\left( {\theta ,r} \right){\mathbf{h}^H}\left( {\theta ,r} \right)} d\theta dr,
\end{align}
and
\begin{align}
\hat{\mathbf{R}}_{e}=\mathop{{\int\!\!\!\!\!\int}\mkern-21mu \bigcirc}\limits_{Area_{e}}
 {\mathbf{h}\left( {\theta ,r} \right){\mathbf{h}^H}\left( {\theta ,r} \right)} d\theta dr.
\end{align}

To guarantee all users have a good secrecy performance, we maximize the SLNR of the user having the worst performance. Thus, the designing of beamforming vector becomes an optimization problem as follows
\begin{align}\label{min}
\begin{array}{l}
\mathop {{\rm{maximize}}}\limits_\mathbf{v}  ~\mathop {\min }\limits_{m=1,\cdots,M} \left\{\hat{\mathrm{SLNR}}_m\right\},\\
\mathrm{s.t.}~{\mathbf{v}^H}\mathbf{v} =1 ,
\end{array}
\end{align}
and it can be converted into an equivalent problem given by
\begin{align}\label{OPT}
\begin{array}{l}
\mathop {{\rm{maximize}}}\limits_\mathbf{v} ~\mathop {\min }\limits_{m=1,\cdots,M} \left\{\ln\left(\hat{\mathrm{SLNR}}_m\right)\right\},\\
\mathrm{s.t.}~{\mathbf{v}^H}\mathbf{v} = 1.
\end{array}
\end{align}

Next, to transform the problem into a convex problem, first, we substitute the numerator and denominator of the fraction in the objective function by exponential variables
\begin{align}\label{a}
{e^{{a_m}}} = {\alpha _1}P\text{tr}\left\{ {\mathbf{R}_v{\hat{\mathbf{R}}_{{d_m}}}} \right\},
\end{align}
and
\begin{align}\label{b}
{e^{{b}}} = \text{tr}\left\{ {\mathbf{R}_v\left( {{\alpha _1}P{\hat{\mathbf{R}}_e} + \sigma _e^2{I_N}} \right)} \right\},
\end{align}
with
\begin{align}
\mathbf{R}_v=\mathbf{v}\mathbf{v}^H.
\end{align}
Then, according to the basic properties of the exponential and logarithmic functions, the objective function in (\ref{OPT}) can be rewritten as
\begin{align}
\mathop {\min }\limits_{m=1,\cdots,M} \left\{a_m-b\right\}.
\end{align}
Meanwhile, $a_m$ and $b$ are constrained by the right hand sides of equation in (\ref{a}) and (\ref{b}).
Accordingly, the optimization problem can be transformed as
\begin{subequations}
\begin{align}
&\mathop {{\rm{maximize}}}\limits_\mathbb{S} ~\mathop {\min }\limits_{m=1,\cdots,M} \left\{a_m-b\right\},\\
&\mathrm{s.t.}\nonumber\\
&{\alpha _1}P\text{tr}\left\{ {\mathbf{R}_v{\hat{\mathbf{R}}_{{d_m}}}} \right\}\geq {e^{{a_m}}},\forall m,\\
&\text{tr}\left\{ {\mathbf{R}_v\left( {{\alpha _1}P{\hat{\mathbf{R}}_e} + \sigma^2{I_N}} \right)} \right\}\leq e^{{b}},\label{24}\\
&\text{tr}\left(\mathbf{R}_v\right)=1,\\
&\mathbf{R}_v\succeq \mathbf{0}.
\end{align}
\end{subequations}

Here, the set $\mathbb{S}$ refers to $\{a_1,\ldots, a_m, b, \mathbf{R}_v\}$. And $v$ is obtained by the decomposition of $\tilde{R}_v[n]=vv^H$ in the case of rank$(\tilde{R}_v[n])=1$, otherwise the randomization technology \cite{zhang2013} would be utilized to get a rank-one approximation. It can be observed that the objective function is a concave function but the constraint (\ref{24}) is non-convex. Therefore, we linearize $e^b$ based on the first-order Taylor approximation to transform the constraint into a convex one as follows
\begin{align}
e^b=e^{\bar b}(b-\bar b+1),
\end{align}
where
\begin{align}\label{bar_b}
\bar b=\ln\left(\text{tr}\left\{ {\mathbf{R}_v\left( {{\alpha _1}P{\hat{\mathbf{R}}_e} + \sigma _e^2{I_N}} \right)} \right\}\right),
\end{align}
is the point around which the approximation are made. Thus the problem becomes
\begin{subequations}\label{c}
\begin{align}
&\mathop {{\rm{maximize}}}\limits_\mathbb{S} ~\mathop {\min }\limits_{m=1,\cdots,M} \left\{a_m-b\right\}\\
&\mathrm{s.t.}\nonumber\\
&{\alpha _1}P\text{tr}\left\{ {\mathbf{R}_v{\hat{\mathbf{R}}_{{d_m}}}} \right\}\geq {e^{{a_m}}},\forall m,\\
&\text{tr}\left\{ {\mathbf{R}_v\left( {{\alpha _1}P{\hat{\mathbf{R}}_e} + \sigma _e^2{I_N}} \right)} \right\}\leq e^{\bar b}\left(b-\bar b+1\right),\\
&\text{tr}\left(\mathbf{R}_v\right)=1,\\
&\mathbf{R}_v\succeq \mathbf{0}.
\end{align}
\end{subequations}
Clearly, the original problem (\ref{min}) is approximately transformed into an equivalent convex problem \cite{Boyd2004Convex}, it can be solved iteratively and converges by Algorithm 1 using optimization software such as CVX. It is noteworthy that $\mathbf{R}_v$ has to satisfy rank($\mathbf{R}_v$)=1. Otherwise the randomization technology would be utilized to get a rank-one approximation.

\begin{algorithm}[t]
\caption{Algorithm for solving problem (\ref{c})}
\label{alg:A}
\begin{algorithmic}[1]
\STATE {Given $\tilde{\mathbf{v}}$ randomly that is feasible to (\ref{c});}
\STATE {Set $\tilde{R}_v[0]=\tilde{\mathbf{v}}\tilde{\mathbf{v}}^H$ and set $n=0$;}
\REPEAT
\STATE Substituting $\tilde{R}_v[n]$ into (\ref{bar_b}) yields $\bar b[n+1]$;
\STATE updating $n=n+1$;
\STATE Substituting $\bar b[n]$ into (\ref{c}) yields the optimal solution $\tilde{R}_v[n]$;
\UNTIL{Convergence}
\STATE Obtain $\mathbf{v}$ by the decomposition of $\tilde{R}_v[n]=\mathbf{v}\mathbf{v}^H$ in the case of rank$(\tilde{R}_v[n])=1$; otherwise the randomization technology would be utilized to get a rank-one approximation.
\end{algorithmic}
\end{algorithm}

Based upon the feasible solution of the beamforming vector $\mathbf{v}$, due to the fact that the expression of SLNR does not involve $\mathbf{w}$ in the next step, the AN matrix can be obtained by maximizing the minimal ANLNR
The optimization problem for $\mathbf{w}$ can be write as
\begin{align}
&\mathop {{\rm{maximize}}}\limits_{\mathbf{w}} ~\mathop {\min }\limits_{m=1,\cdots,M} \left\{\hat{\mathrm{ANLNR}}_m\right\},\nonumber\\
&\mathrm{s.t.}~\mathbf{w}^H\mathbf{w}=1.
\end{align}

Similar to Eq. (\ref{c}), the optimization problem can be transformed as
\begin{subequations}\label{c_w}
\begin{align}
&\mathop {{\rm{maximize}}}\limits_\mathbb{S} ~\mathop {\min }\limits_{m=1,\cdots,M} \left\{c-d_m\right\}\\
&\mathrm{s.t.}\nonumber\\
&{\alpha _1}P\text{tr}\left\{ {\mathbf{R}_w{\hat{\mathbf{R}}_{{d_m}}}} \right\}\geq {e^{{c}}},\\
&\text{tr}\left\{ {\mathbf{R}_w\left( {{\alpha _1}P{\hat{\mathbf{R}}_e} + \sigma _e^2{I_N}} \right)} \right\}\leq e^{\bar d_m}\left(d_m-\bar d_m+1\right),\forall m,\\
&\text{tr}\left(\mathbf{R}_w\right)=1,\\
&\mathbf{R}_w\succeq \mathbf{0}.
\end{align}
\end{subequations}
where $R_w=\mathbf{w}\mathbf{w}^H$, and
\begin{align}\label{bar_b_m}
\bar d_m=\ln\left(\text{tr}\left\{ {\mathbf{R}_w\left( {{\alpha _1}P{\hat{\mathbf{R}}_e} + \sigma _e^2{I_N}} \right)} \right\}\right),
\end{align}
Hence, we can obtain the optimal $\mathbf{w}$ by Algorithm 2.
Since problem (\ref{c}) and (\ref{c_w}) both have one $N_T \times N_T$ matrix variable, these methods need at most $O((N_T)^{3.5}log(1/\epsilon))$ calculations at each inner iteration \cite{Ouyang2017Secrecy}. As ${\hat{\mathbf{R}}_{{d_m}}}$ is calculated by a integration, which needs $O(\frac{r_{max}\theta_{max}}{\Delta r \Delta \theta}(2N_T-1))$ calculations, where $\Delta \theta$ and $\Delta r$ are the integral precision, and generally they are set to be very small. We define $N_p=\frac{r_{max}\theta_{max}}{\Delta r \Delta \theta}$ as the integral points. Thus the per iteration complexity of the proposed R-SLNR-ANLNR maximization based scheme can be approximately calculated as  $O(2((N_T)^{3.5}log(1/\epsilon)+N_p(2N_T-1)))$.
\begin{algorithm}[t]
\caption{Algorithm for solving problem (\ref{c_w})}
\label{alg:A}
\begin{algorithmic}[1]
\STATE {Given $\tilde{\mathbf{w}}$ randomly that is feasible to (\ref{c_w});}
\STATE {Set $\tilde{R}_w[0]=\tilde{\mathbf{w}}\tilde{\mathbf{w}}^H$ and set $n=0$;}
\REPEAT
\STATE Substituting $\tilde{R}_w[n]$ into (\ref{bar_b_m}) yields $\bar d_m[n+1]$;
\STATE updating $n=n+1$;
\STATE Substituting $\bar b[n]$ into (\ref{c_w}) yields the optimal solution $\tilde{R}_w[n]$;
\UNTIL{Convergence}
\STATE Obtain $\mathbf{w}$ by the decomposition of $\tilde{R}_w[n]=\mathbf{w}\mathbf{w}^H$ in the case of rank$(\tilde{R}_w[n])=1$; otherwise the randomization technology would be utilized to get a rank-one approximation.
\end{algorithmic}
\end{algorithm}

\section{Method based on point SLNR maximization}
Although the robust method based on regional SLNR and ANLNR maximization has guaranteed a stable performance for users in the whole error region, a sum of confidential signal energy in the error region has to be calculated, which may bring a huge amount of calculation complexity. To reduce complexity, we propose a scheme based on maximizing the point secrecy capacity which is capable of significantly reducing the algorithm complexity with comparable performance.

First, we define the width of the main-lobe in the angle and distance dimension as \cite{Sadek2007A},
\begin{align}
{\Theta_{m}} = \left[ {{\theta _{{d_m}}} - \frac{{{\theta _{BW}}}}{2},{\theta _{{d_m}}} + \frac{{{\theta _{BW}}}}{2}} \right],
\end{align}
and
\begin{align}
{{D}_m} = \left[ {{r_{{d_m}}} - \frac{c}{B},{r_{{d_m}}} + \frac{c}{B}} \right],
\end{align}
where
\begin{align}
{\theta _{BW}} = \frac{{2\lambda }}{{Nd}}.
\end{align}
Clearly, it is reasonable that the estimate error range is smaller than the main-beam range. Consider the scenario with 16 antennas and 5MHz total bandwidth, the conditions that
\begin{align}
\frac{\lambda }{{Nd}} = \frac{1}{{8}} > \Delta {\theta _{\max }},
\frac{c}{B} = \frac{{3 \times {{10}^8}}}{{5 \times {{10}^6}}} = 60 > \Delta {r_{\max }},
\end{align}
with $d=\lambda /2$ is easy to be satisfied. Thus we assume a scenario where the estimate error of the direction angle and the distance is small enough that we have
\begin{align}
Area_{d_m}\subset Area_{MB}(\theta,r)~,~(\theta,r)\in Area_{d_m},
\end{align}
where $Area_{MB}(\theta,r)$ is the main-lobe region, and this means the main-beam always contains the estimate error region. In this case, the optimization point in the estimation error region occupies most of the main-lobe, so as to improve the performance.
Next, we find several points in estimate regions of each user as desired points and several points in wiretap region as eavesdrop points. Then maximize the secrecy performance in those points positions. In this context, we choose four points for instance as follows
\begin{align}
\begin{array}{*{20}{c}}
{p_{dm}^1 = (\hat \theta _{_{{s_{dm}}}}^1,\hat r_{_{{s_{dm}}}}^1) = ({{\hat \theta }_{{d_m}}} - \Delta {\theta _{{\rm{max}}}},{{\hat r}_{{d_m}}}),}\\
{p_{dm}^2 = (\hat \theta _{_{{s_{dm}}}}^2,\hat r_{_{{s_{dm}}}}^2) = ({{\hat \theta }_{{d_m}}} + \Delta {\theta _{{\rm{max}}}},{{\hat r}_{{d_m}}}),}\\
{p_{dm}^3 = (\hat \theta _{_{{s_{dm}}}}^3,\hat r_{_{{s_{dm}}}}^3) = ({{\hat \theta }_{{d_m}}},{{\hat r}_{{d_m}}} - \Delta {r_{{\rm{max}}}}),}\\
{p_{dm}^4 = (\hat \theta _{_{{s_{dm}}}}^4,\hat r_{_{{s_{dm}}}}^4) = ({{\hat \theta }_{{d_m}}},{{\hat r}_{{d_m}}} + \Delta {r_{{\rm{max}}}})},
\end{array}
\end{align}
and

\begin{align}
\begin{array}{*{20}{c}}
{p_{em}^1 = ({{ \theta }_{{{{\mathrm{sidelobe}}_m}}}^1} ,{{ r}_{{{{\mathrm{sidelobe}}_m}}}^1}),}\\
{p_{em}^2 = ({{ \theta }_{{{{\mathrm{sidelobe}}_m}}}^2} ,{{ r}_{{{{\mathrm{sidelobe}}_m}}}^2}),}\\
{p_{em}^3 = ({{ \theta }_{{{{\mathrm{sidelobe}}_m}}}^1},{{ r}_{{{{\mathrm{sidelobe}}_m}}}^1} ),}\\
{p_{em}^4 = ({{ \theta }_{{{{\mathrm{sidelobe}}_m}}}^2},{{ r}_{{{{\mathrm{sidelobe}}_m}}}^2} )},
\end{array}
\end{align}

where

\begin{align}\label{sidelobe}
&{\theta }_{{{{\mathrm{sidelobe}}_m}}}^1=\arccos \left( \cos{{\hat \theta }_{{d_m}}}-\frac{3}{N_T}  \right),\nonumber\\
&{\theta }_{{{{\mathrm{sidelobe}}_m}}}^2=\arccos \left( \cos{{\hat \theta }_{{d_m}}}+\frac{3}{N_T}  \right),\nonumber\\
&{ r}_{{{{\mathrm{sidelobe}}_m}}}^1={{\hat r}_{{d_m}}}-\frac{3c}{2B},\nonumber\\
&{ r}_{{{{\mathrm{sidelobe}}_m}}}^2={{\hat r}_{{d_m}}}+\frac{3c}{2B},
\end{align}
are the maximum values corresponding to the first side-lobes along direction angle and distance dimensions \cite{Shu2018SPWT}.
Then we maximize the SLNR with these points instead of all the region.

To preserve the user with the worst performance, we have the following optimization problem by maximizing the minimal SLNR of the users, given by
\begin{align}\label{O_p}
\begin{array}{l}
\mathop {{\rm{maximize}}}\limits_\mathbf{v}  ~\mathop {\min }\limits_{m=1,\cdots,M} \left\{\left(\hat{\mathrm{SLNR}}_{s_m}\right)\right\},\\
\mathrm{s.t.}~~~{\mathbf{v}^H}\mathbf{v} = 1,
\end{array}
\end{align}
where
\begin{align}
\hat{\mathrm{SLNR}}_{s_m}=\frac{{{\alpha _1}P\text{tr}\left\{ {{{\mathbf{v}}^H}\hat{\mathbf{R}}_{s_m}{{\mathbf{v}}}} \right\}}}{{\text{tr}\left\{ {{{\mathbf{v}}^H}\left( {{\alpha _1}P\hat{\mathbf{R}}_e + \sigma _e^2{\mathbf{I}_N}} \right){{\mathbf{v}}}} \right\}}},
\end{align}
and
\begin{align}
\hat{\mathbf{R}}_{s_m}=\sum\limits_{i = 1}^4 {\mathbf{h}\left( {\theta _{{s_m}}^i,r_{{s_m}}^i} \right){\mathbf{h}^H}\left( {\theta _{{s_m}}^i,r_{{s_m}}^i} \right)}.
\end{align}

\begin{align}\label{Re}
\hat{\mathbf{R}}_{e}=\sum\limits_{m = 1}^M\sum\limits_{i = 1}^4 {\mathbf{h}\left( {\theta _{{e_m}}^i,r_{{e_m}}^i} \right){\mathbf{h}^H}\left( {\theta _{{e_m}}^i,r_{{e_m}}^i} \right)}.
\end{align}

Compared to (\ref{R_d_m}), $\hat{\mathbf{R}}_{s_m}$ is only a sum of several points instead of the integration in all error regions which significantly reduces the calculation and complexity. The following steps to solve the optimal $\mathbf{R}_v$ and $\mathbf{R}_w$ that satisfy rank($\mathbf{R}_v$)=1 and rank($\mathbf{R}_w$)=1, are similar as (\ref{min}) in Section III. To elaborate, we first transform the optimization problem (\ref{O_p}) into an equivalent problem as
\begin{align}
\begin{array}{l}
\mathop {{\rm{maximize}}}\limits_\mathbf{v} ~\mathop {\min }\limits_{m=1,\cdots,M} \left\{\mathrm{ln}\left(\hat{\mathrm{SLNR}}_{s_m}\right)\right\},\\
\mathrm{s.t.}~{\mathbf{v}^H}\mathbf{v} = 1.
\end{array}
\end{align}

Next, we substitute both the numerators and denominators of the fraction in the objective function by exponential variables as
\begin{align}\label{a1}
{e^{{a_m}}} = {\alpha _1}P\text{tr}\left\{ {\mathbf{R}_v{\hat{\mathbf{R}}_{{s_m}}}} \right\},
\end{align}
\begin{align}\label{b1}
{e^{{b}}} = \text{tr}\left\{ {\mathbf{R}_v\left( {{\alpha _1}P{\hat{\mathbf{R}}_e} + \sigma _e^2{I_N}} \right)} \right\}.
\end{align}
Then, we expend $e^b$ by the first-order Taylor approximation as
\begin{align}
e^b=e^{\bar b}(b-\bar b+1).
\end{align}
Consequently, the final optimization problem becomes a convex optimization problem, given by
\begin{subequations}\label{c1}
\begin{align}
&\mathop {{\rm{maximize}}}\limits_\mathbb{S} ~\mathop {\min }\limits_{m=1,\cdots,M} \left\{a_m-b\right\},\\
&\mathrm{s.t.}\nonumber\\
&{\alpha _1}P\text{tr}\left\{ {\mathbf{R}_v{\hat{\mathbf{R}}_{{s_m}}}} \right\}\geq {e^{{a_m}}},\forall m,\\
&\text{tr}\left\{ {\mathbf{R}_v\left( {{\alpha _1}P{\hat{\mathbf{R}}_e} + \sigma _e^2{I_N}} \right)} \right\}\leq e^{\bar b}(b-\bar b+1),\\
&\text{tr}(\mathbf{R}_v)=1,\\
&\mathbf{R}_v\succeq \mathbf{0},
\end{align}
\end{subequations}
where
\begin{align}\label{bar_d}
\bar b=\ln\left(\text{tr}\left\{ {\mathbf{R}_v\left( {{\alpha _1}P{\hat{\mathbf{R}}_e} + \sigma _e^2{I_N}} \right)} \right\}\right).
\end{align}

\begin{figure}[t]
\centering
\includegraphics[width=0.55\textwidth]{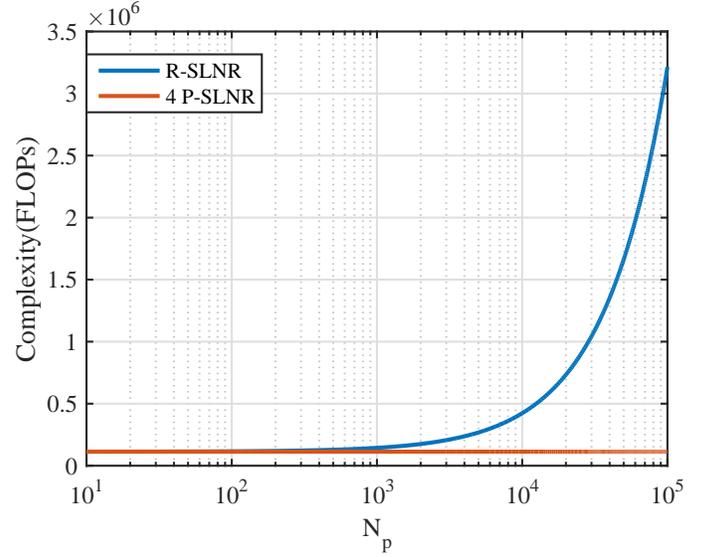}\\
\caption{The complexity versus the integral points $N_p$ of the proposed methods.}\label{complexity.eps}
\end{figure}

Then, we can obtain the optimal $R_v$ with Algorithm 1 by optimization software. Note that $\mathbf{R}_v$ has to satisfy rank($\mathbf{R}_v$)=1, otherwise the randomization technology would be utilized to get a rank-one approximation.


\begin{figure}[t]
\centering
\subfigure[R-SLNR]{
\label{leakage.eps}
\includegraphics[width=0.48\textwidth]{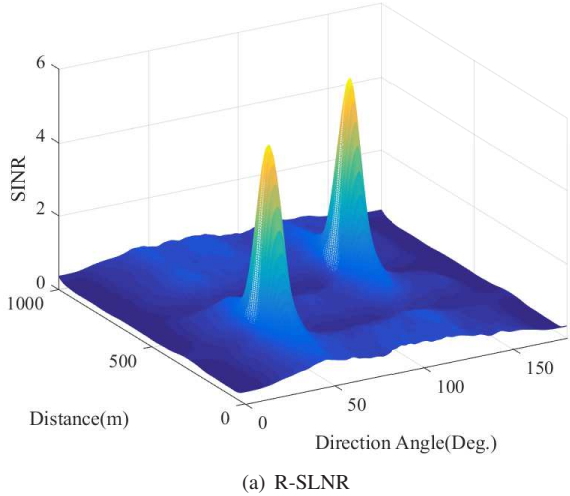}}
\hspace{1in}
\subfigure[P-SLNR]{
\label{point_leakage.eps}
\includegraphics[width=0.48\textwidth]{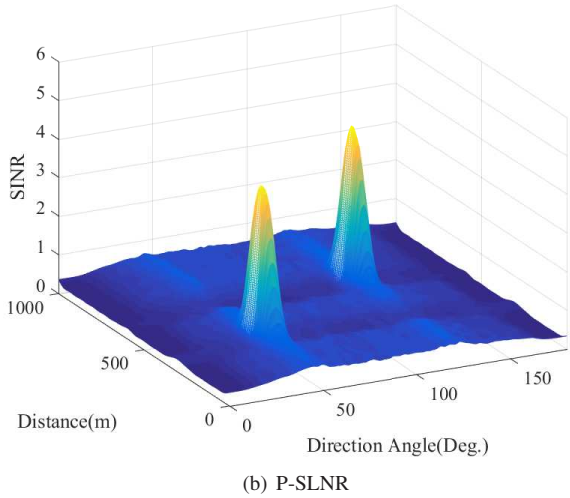}}
\caption{The 3-D surface of SINR versus the direction angle and the distance of the proposed methods.}
\label{SINR}
\end{figure}
Similarly, the AN matrix can be obtained by maximizing the ANLNR at the desired position with following optimization problem.
\begin{subequations}\label{c2_w}
\begin{align}
&\mathop {{\rm{maximize}}}\limits_\mathbb{S} ~\mathop {\min }\limits_{m=1,\cdots,M} \left\{c-d_m\right\}\\
&\mathrm{s.t.}\nonumber\\
&{\alpha _1}P\text{tr}\left\{ {\mathbf{R}_w{\hat{\mathbf{R}}_{{s_m}}}} \right\}\geq {e^{{c}}},\\
&\text{tr}\left\{ {\mathbf{R}_w\left( {{\alpha _1}P{\hat{\mathbf{R}}_e} + \sigma _e^2{I_N}} \right)} \right\}\leq e^{\bar d_m}\left(d_m-\bar d_m+1\right),\forall m,\\
&\text{tr}\left(\mathbf{R}_w\right)=1,\\
&\mathbf{R}_w\succeq \mathbf{0}.
\end{align}
\end{subequations}
Accordingly, we can obtain the optimal $\mathbf{w}$ with Algorithm 2.

As ${\hat{\mathbf{R}}_{{s_m}}}$ is calculated by a summation, which needs $O(M(2N_T-1))$ calculations, where $M$ is the number of point. Thus, the per iteration complexity of the proposed P-SLNR-ANLNR maximization based scheme can be approximately calculated as  $O(2((N_T)^{3.5}log(1/\epsilon)+M(2N_T-1)))$. Compared to Algorithm 1, the complexity significantly reduces since we have $N_p\gg M$, and the complexity superiority will increase with the improve of the integral precision $\Delta r$ and $\Delta \theta$ which are usually tend to zero. Moreover, the smaller the selected sampling points are, the lower the method complexity will be.
Fig. \ref{complexity.eps} shows the complexity versus the integral points $N_p$ of the R-SLNR-ANLNR and P-SLNR-ANLNR schemes with the number of antenna $N_T=16$. We can observe that the complexity of the R-SLNR-ANLNR maximization-based scheme increases a lot with the integral points grows, but of the P-SLNR-ANLNR maximization-based scheme is constant. In practice, the integral precision $\Delta \theta$ and $\Delta r$ are very small, and thus the integral points tend to be extremely large which lead the R-SLNR-ANLNR maximization-based scheme has a much larger complexity than P-SLNR-ANLNR maximization-based scheme.

\begin{figure}[t]
\centering
\subfigure[R-SLNR]{
\label{leakage_topview.eps}
\includegraphics[width=0.48\textwidth]{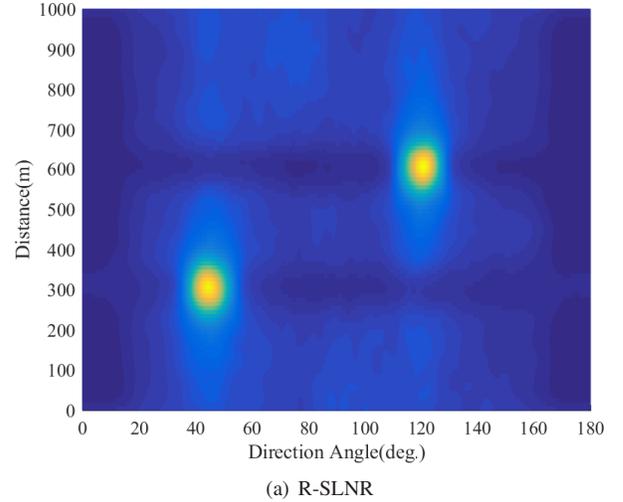}}
\hspace{1in}
\subfigure[P-SLNR]{
\label{point_leakage_topview.eps}
\includegraphics[width=0.48\textwidth]{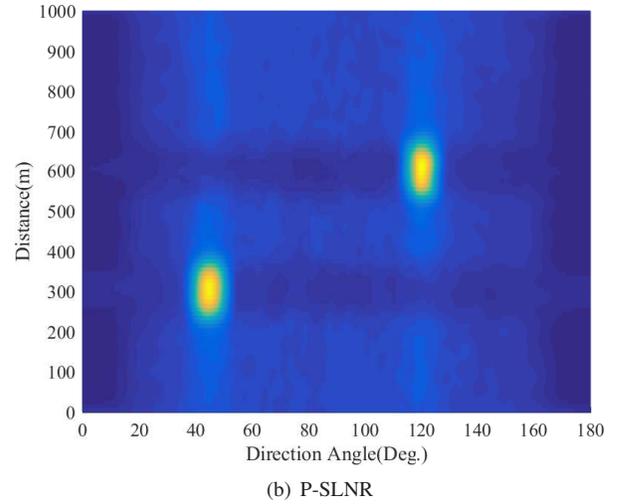}}
\caption{The top view of SINR versus the direction angle and the distance of the proposed methods.}
\label{SINR topview}
\end{figure}

\section{simulation results and analysis}
In this section, we evaluate the performance of our proposed regional robust SPWT schemes by numerical simulations. The default system parameters are chosen as shown in Table \ref{tab}, the position coordinates are referenced by the transmitter, and the measurement errors are modelled as independently identically distributed (i.i.d) random variables and distributed in truncated Gaussian. In this part, the secrecy rate (SR) is defined as
\begin{align}\label{SR}
\mathrm{SR}=\min \limits_{m=1,\cdots,M}\left[\min \limits_{k=1,\cdots,K}\left(C_{d_m}-C_{e_k}\right) \right],
\end{align}
where $C_{d_m}=\log_2(1+\mathrm{SINR}_{d_m})$, and $C_{e_k}=\log_2(1+\mathrm{SINR}_{e_k})$.

Firstly, Fig. \ref{leakage.eps} and Fig. \ref{point_leakage.eps} illustrates the 3-D performance surface of SINR versus direction angle $\theta$ and distance $R$ of the proposed method. It can be observed that there are only two high signal energy peaks of confidential messages formed around the two desired positions. Outside the main peak or the estimation error region, the SINR is far lower than that of the two desired users. By our coarse measurement, the average SINR outside the estimation error region is only one tenth of that in the two desired users.
Then, Fig. \ref{leakage_topview.eps} and Fig. \ref{point_leakage_topview.eps} illustrates the top view of Fig. \ref{leakage.eps} and Fig. \ref{point_leakage.eps}. We can clearly observe that the confidential energy concentrates on the estimation error region around the two users. From Fig. \ref{SINR}, Fig. \ref{SINR topview} and Fig. \ref{complexity.eps}, we observe that P-SLNR-ANLNR maximization-based scheme has a comparable secrecy performance to the R-SLNR-ANLNR maximization-based scheme while its complexity has significantly decreased in practice.
Compared to the conventional direction modulation, which yield a high energy ridge or mountain chain in direction and distance dimensions, our proposed SPWT schemes can converge the confidential signal energy into a point and yield a energy peak around the point. Thus SPWT has a higher energy efficiency.
%

\begin{table}[t]
\centering
\caption{SIMULATION PARAMETERS SETTING}\label{tab}
 \begin{tabular}{|c|c|}
\hline
Parameter&Value\\
\hline
The number of antennas at the transmitter ($N_T$)&16 \\
\hline
The number of desired users (M)&2\\
\hline
Total signal bandwidth (B)&5MHz\\
\hline
Total transmit power ($P$)&1W \\
\hline
Power allocation factors (\{$\alpha_1^2$,$\alpha_2^2$\})&\{0.5,0.5\}\\
\hline
\tabincell{c}{Desired users positions \\ $({(\theta_{d_1},R_{d_1}),(\theta_{d_2},R_{d_2})})$}&\tabincell{c}{$(45^\circ,300m)$,\\$(120^\circ,600m)$}\\
\hline
Central carrier frequency ($f_c$)&3GHz\\
\hline
Subcarriers number (N)&1024\\
\hline
Signal to noise ratio(SNR)&15dB\\
\hline
\tabincell{c}{The maximal observe range of the angle \\and distance ({$\theta_{max},r_{max}$})}&{$180^{\circ},1000m$}\\
\hline
The maximal estimation error ({$\Delta\theta_{max},\Delta r_{max}$})&{$6^{\circ},50m$}\\
\hline
\end{tabular}
\end{table}

\begin{figure}[t]
\centering
\includegraphics[width=0.480\textwidth]{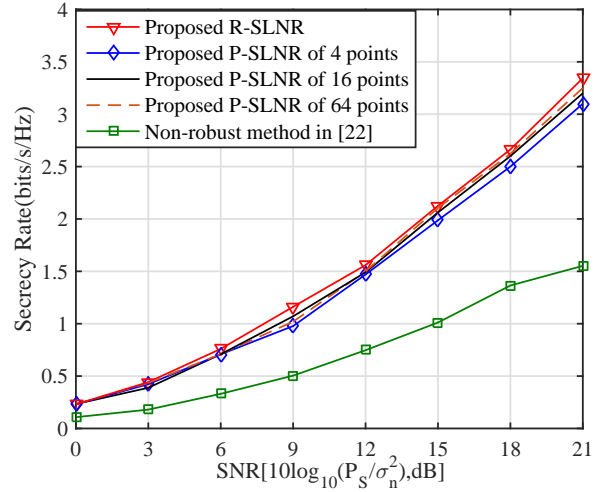}\\
\caption{Secrecy rate versus SNR($10log_{10}(P/\sigma_n^2)$) of the proposed methods with $N_T=16$, $B=5MHz$, $\Delta \theta_{max}=6^{\circ}$ and $\Delta r_{max}=50m$.}\label{SR.eps}
\end{figure}

\begin{figure}[t]
\centering
\subfigure[$N_T=8$]{
\label{N_T=8.eps}
\includegraphics[width=0.48\textwidth]{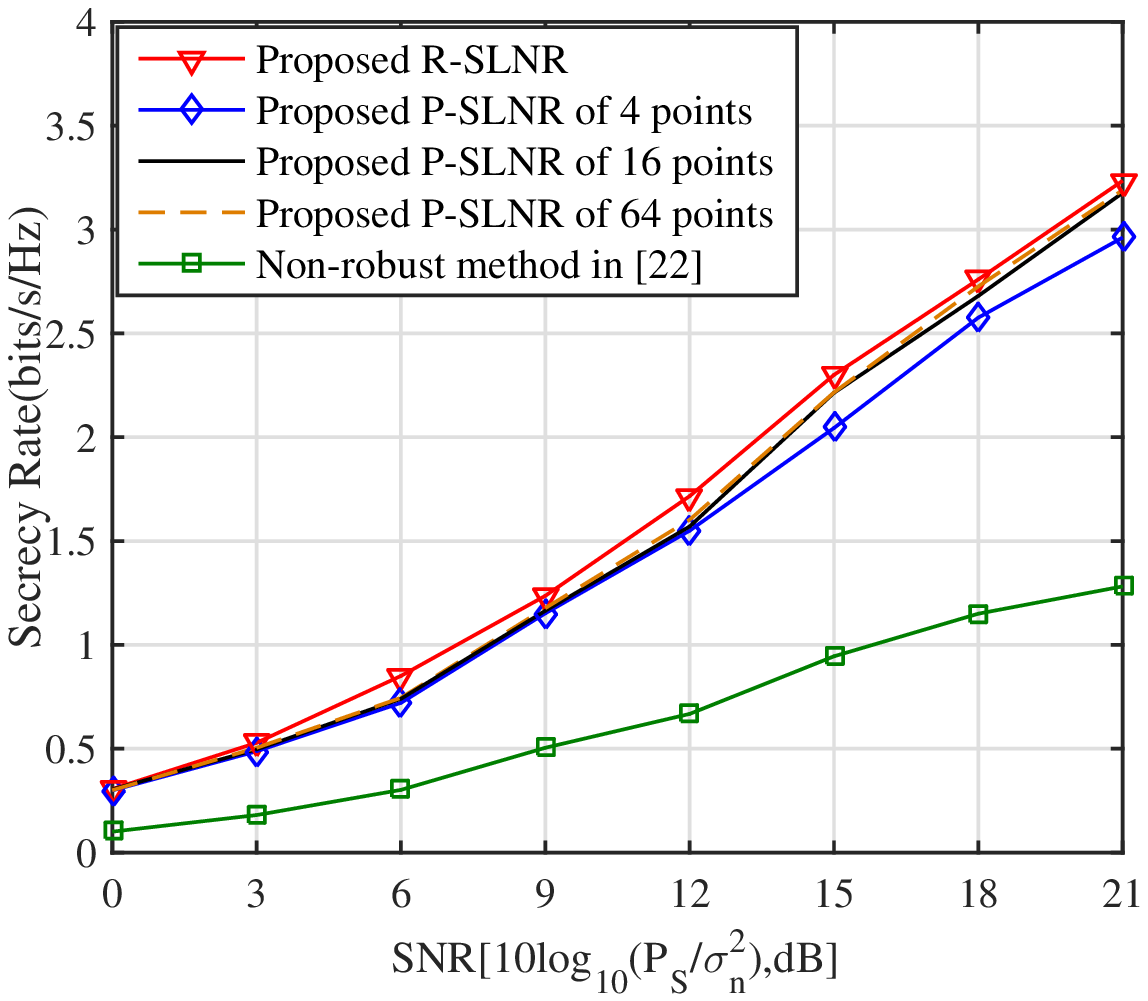}}
\hspace{1in}
\subfigure[$N_T=32$]{
\label{N_T=32.eps}
\includegraphics[width=0.48\textwidth]{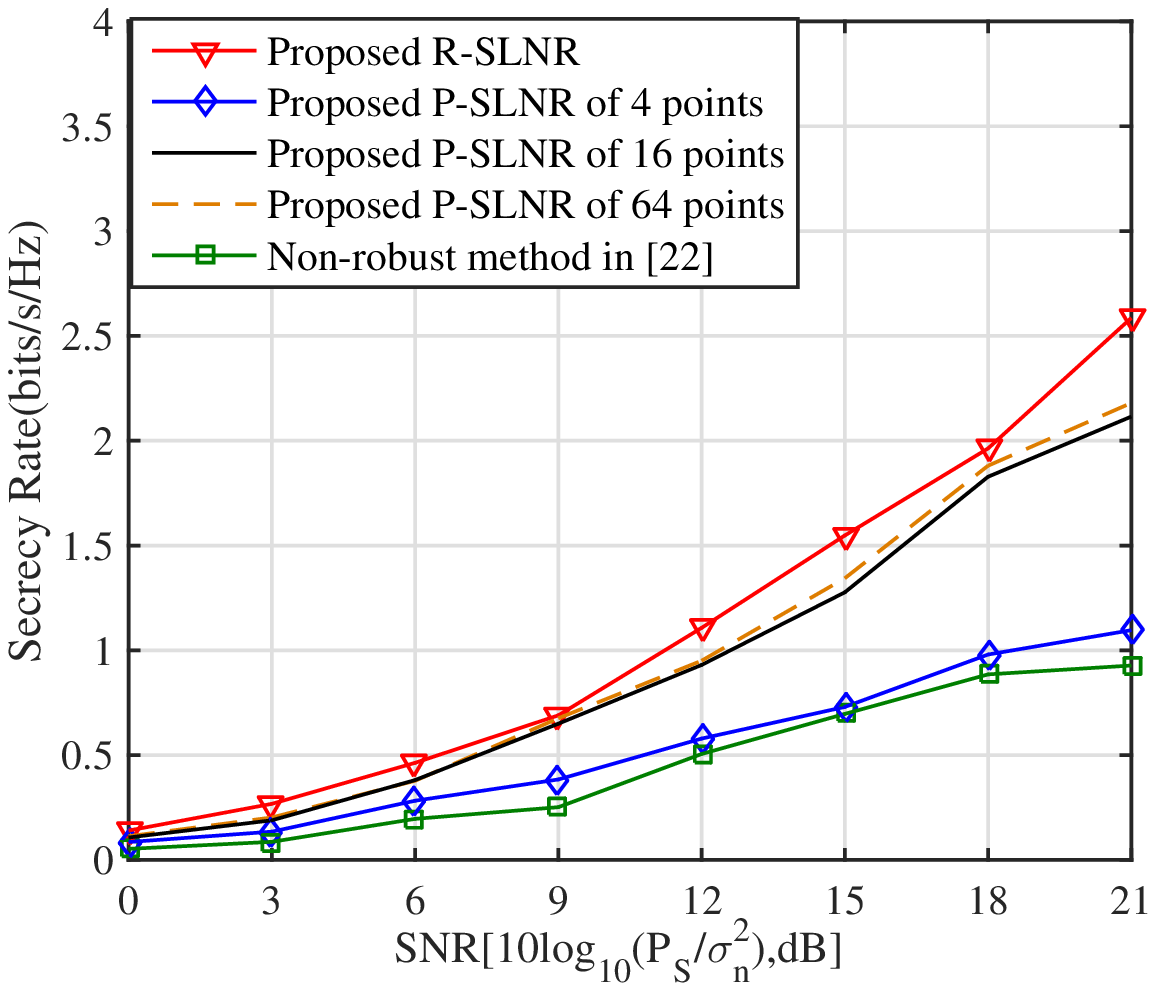}}
\caption{The curves of secrecy rate versus SNR($10log_{10}(P/\sigma_n^2)$) of the proposed methods with $N_T=8$ and $N_T=32$.}
\label{SR1}
\end{figure}
%

%

In Fig. \ref{SR.eps}, the curve of secrecy rates versus SNR for the proposed schemes is plotted. Observing this figure, the secrecy performance of two proposed schemes significantly increases with SNR compared to non-robust scheme. The R-SLNR-ANLNR maximization-based scheme outperforms all the other schemes in terms of the secrecy rate performance, and the secrecy rate of the P-SLNR-ANLNR maximization-based scheme increases as the number of points grows. When $N_T=16$, $B=5MHz$, $\Delta\theta_{max}=6^{\circ}$ and $\Delta r_{max}=50m$, the secrecy rate of the 4,16 and 64 P-SLNR-ANLNR maximization-based scheme are very close. The reason behind this trend is that their main-lobe is large enough compared to the estimation error region. It means that our proposed P-SLNR-ANLNR maximization-based scheme has the comparable performance with a lower complexity. As aforementioned, the width of the main-lobe depends on the antenna number $N_T$ and the total signal bandwidth $B$, and thus we further simulate the secrecy rate of the proposed schemes under different numbers of antennas, signal bandwidth and estimation error range.

Fig. \ref{SR1} draws the performance of the secrecy rate versus SNR under different number of antennas. Compared to Fig. \ref{SR.eps}, we observe the fact that the secrecy rate firstly increases from $N_T=8$ to $N_T=16$, but then decreases from $N_T=16$ to $N_T=32$, and meanwhile the performance of the P-SLNR-ANLNR maximization-based scheme with $4$ points significantly deteriorates with the number of antennas. The reason is as follows: Firstly, when $N_T$ is small, the main-lobe is sufficiently large to guarantee the optimized regions locating in the main-lobe. Then, as the number of antennas increases, the secrecy rate increases from $N_T=8$ to $N_T=16$. Secondly, when $N_T$ grows large, its main-lobe is very small, and then some of the optimized regions run out of the main-lobe, thus decreasing the secrecy performance from $N_T=16$ to $N_T=32$.

\begin{figure}[t]
\centering
\subfigure[B=7.5MHz]{
\label{B=7.5.eps}
\includegraphics[width=0.48\textwidth]{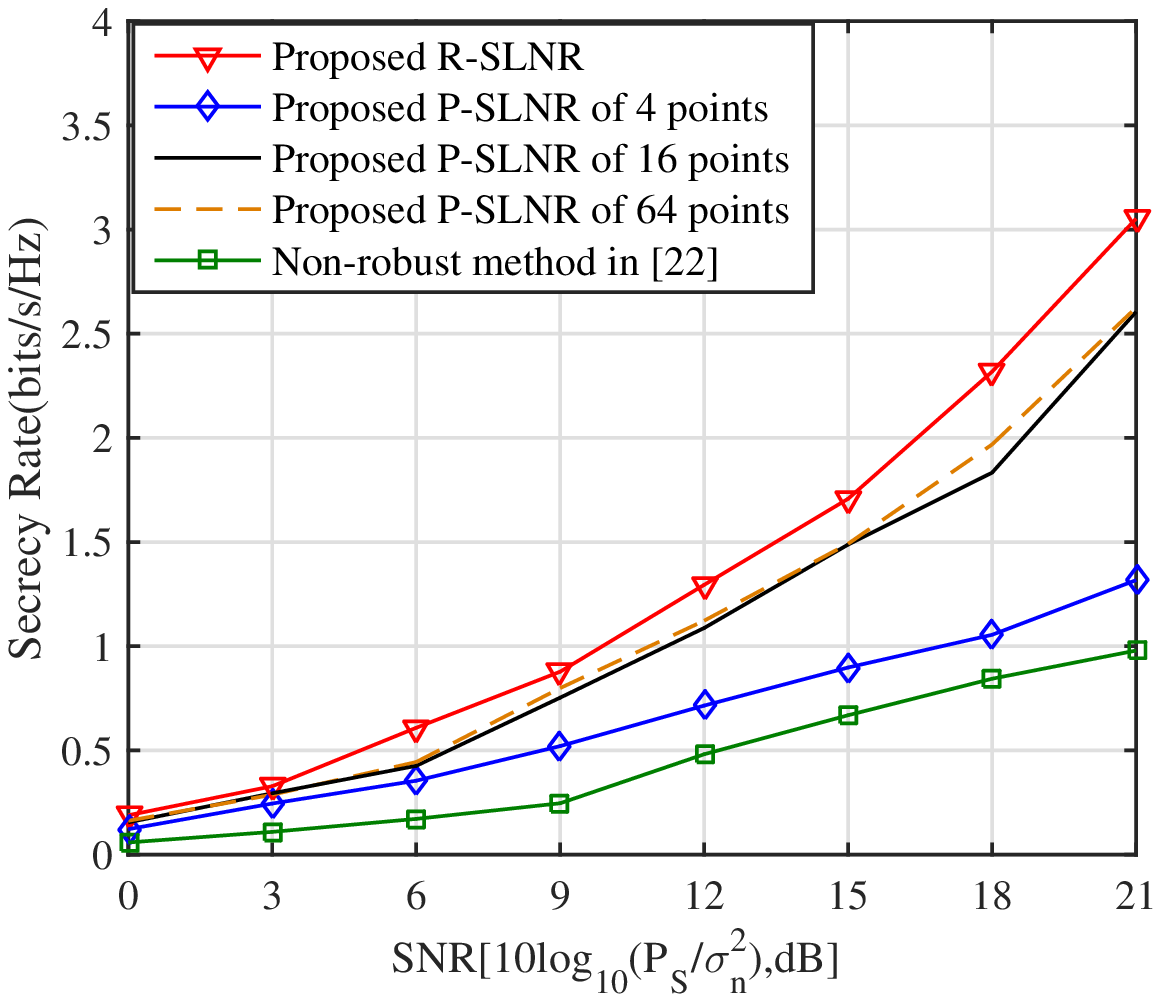}}
\hspace{1in}
\subfigure[B=10MHz]{
\label{B=10.eps}
\includegraphics[width=0.48\textwidth]{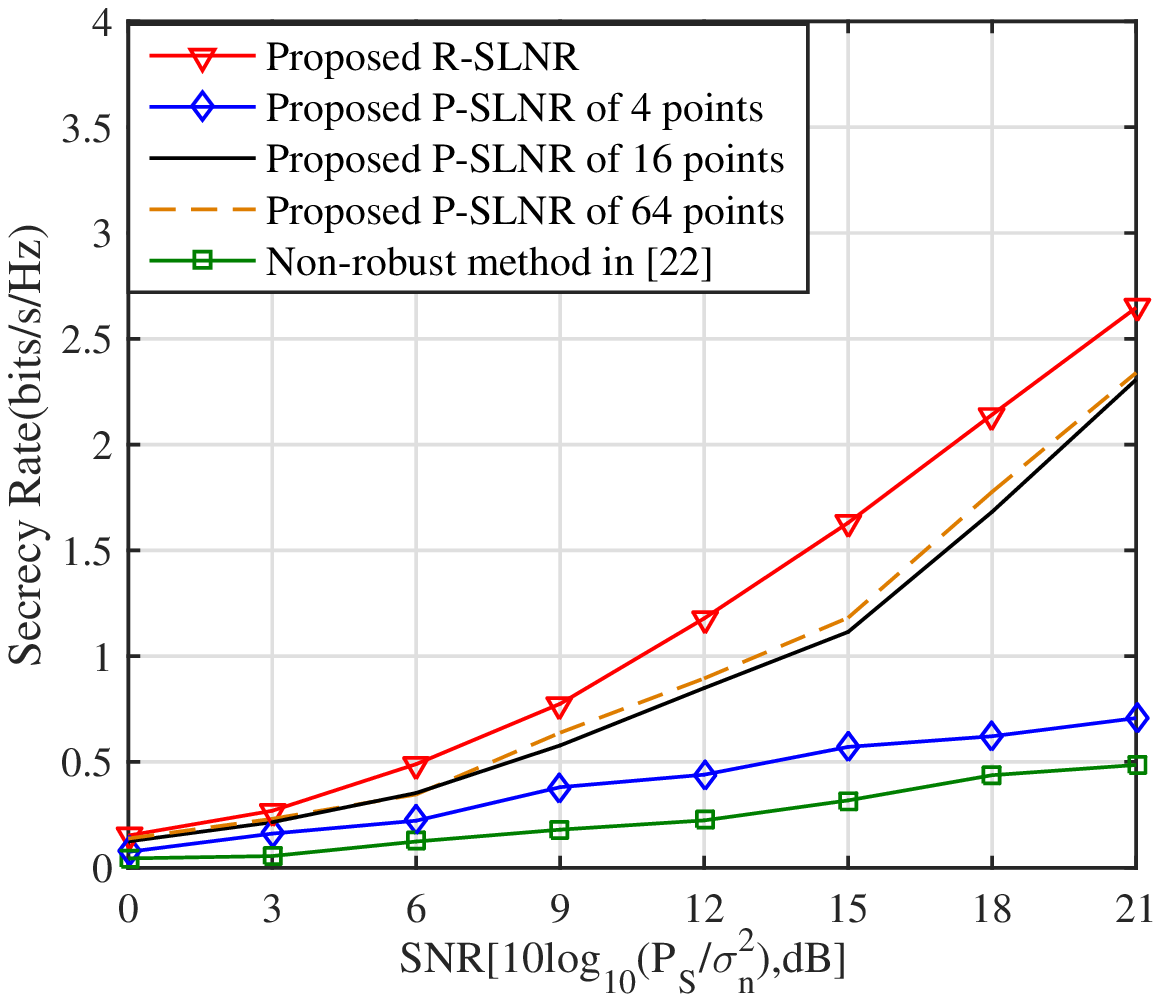}}
\caption{The curves of secrecy rate versus SNR($10log_{10}(P/\sigma_n^2)$) of the proposed methods with $B=7.5MHz$ and $N=10MHz$.}
\label{SR2}
\end{figure}

Fig. \ref{SR2} draws the secrecy rates versus SNR with different signal bandwidth. Compared with Fig. \ref{SR.eps} of $B=5MHz$, it can be observed that with the increase of bandwidth $B$, the total secrecy rate decreases due to the main-lobe become smaller, and the performance of the P-SLNR-ANLNR maximization-based scheme with $4$ points decreases significantly.

\begin{figure}[t]
\centering
\subfigure[$\Delta \theta_{max}=3^{\circ}$]{
\label{angle3.eps}
\includegraphics[width=0.48\textwidth]{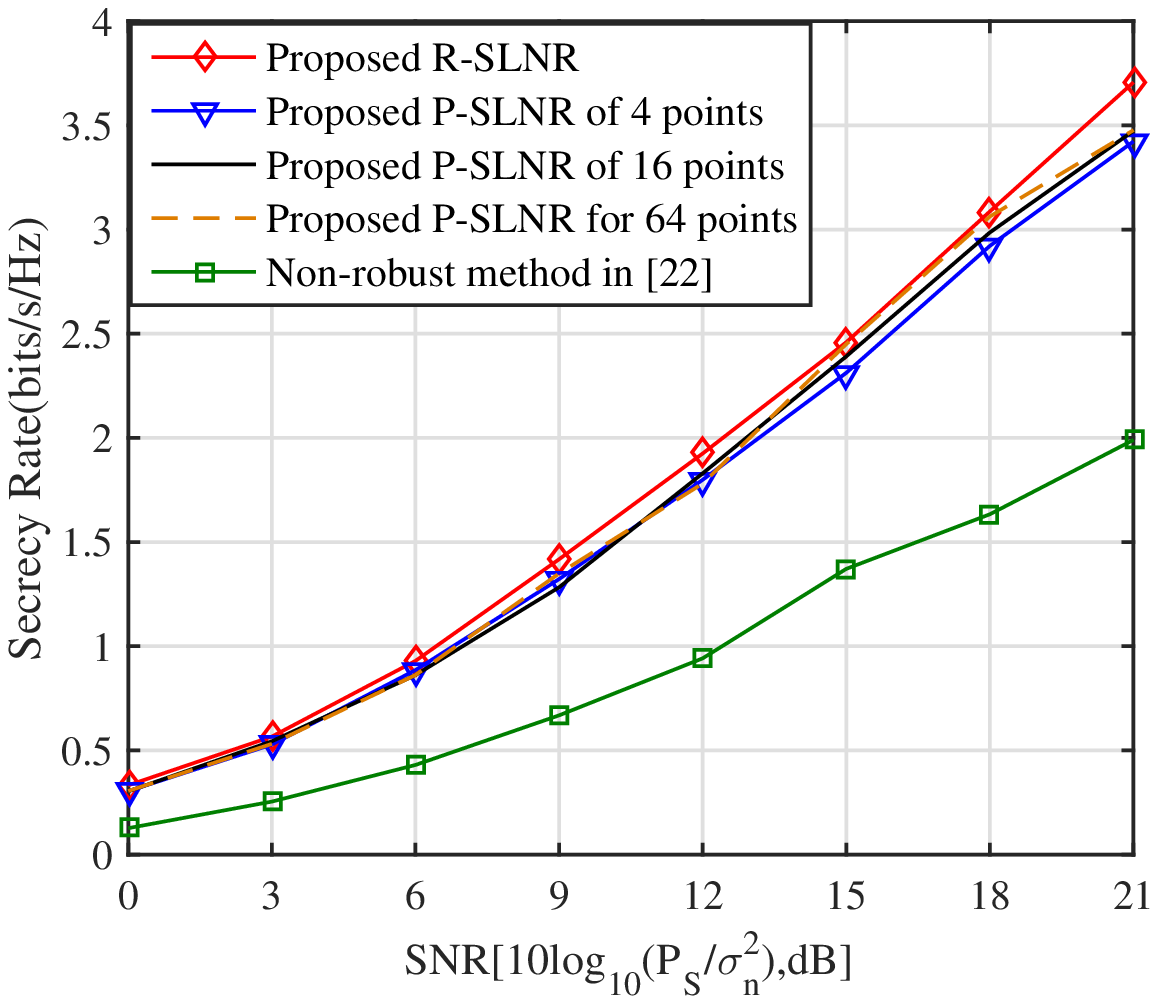}}
\hspace{1in}
\subfigure[$\Delta \theta_{max}=9^{\circ}$]{
\label{angle9.eps}
\includegraphics[width=0.50\textwidth]{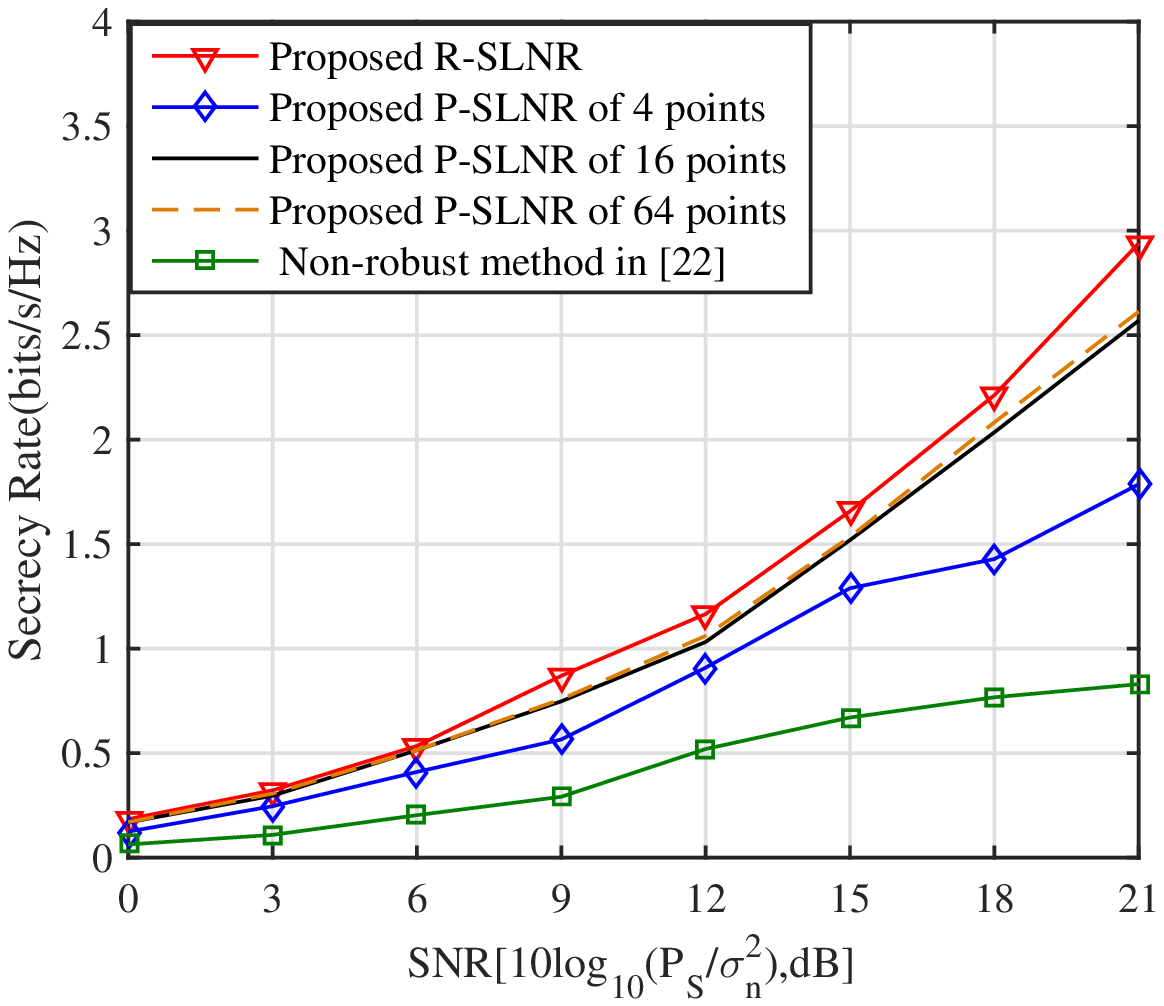}}
\caption{Curves of secrecy rate versus SNR($10log_{10}(P/\sigma_n^2)$) of the proposed methods with $\Delta \theta_{max}=3^{\circ}$ and $\Delta \theta_{max}=9^{\circ}$.}
\label{SR3}
\end{figure}

\begin{figure}[t]
\centering
\subfigure[$\Delta r_{max}=20m$]{
\label{D=20.eps}
\includegraphics[width=0.48\textwidth]{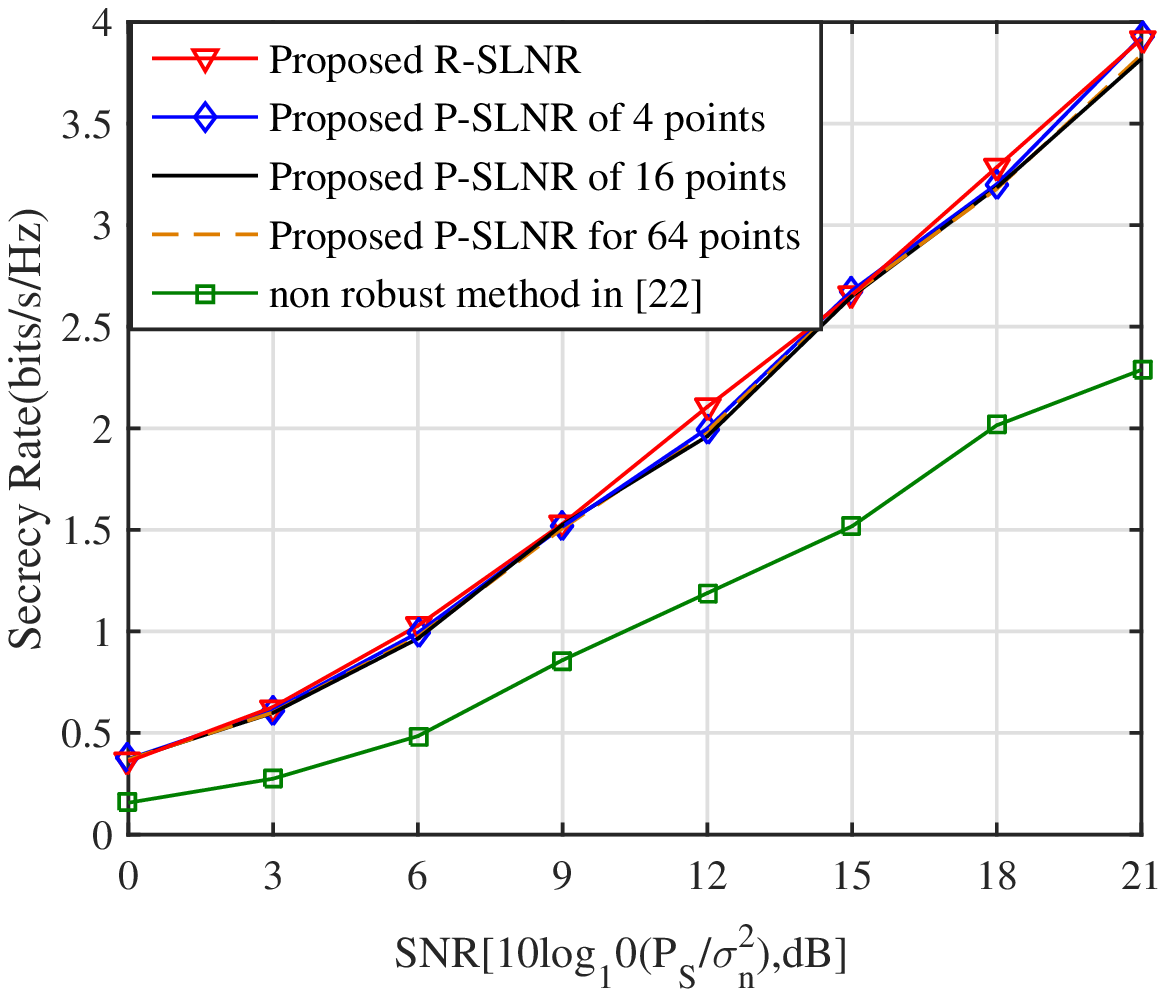}}
\hspace{1in}
\subfigure[$\Delta r_{max}=100m$]{
\label{D=100.eps}
\includegraphics[width=0.50\textwidth]{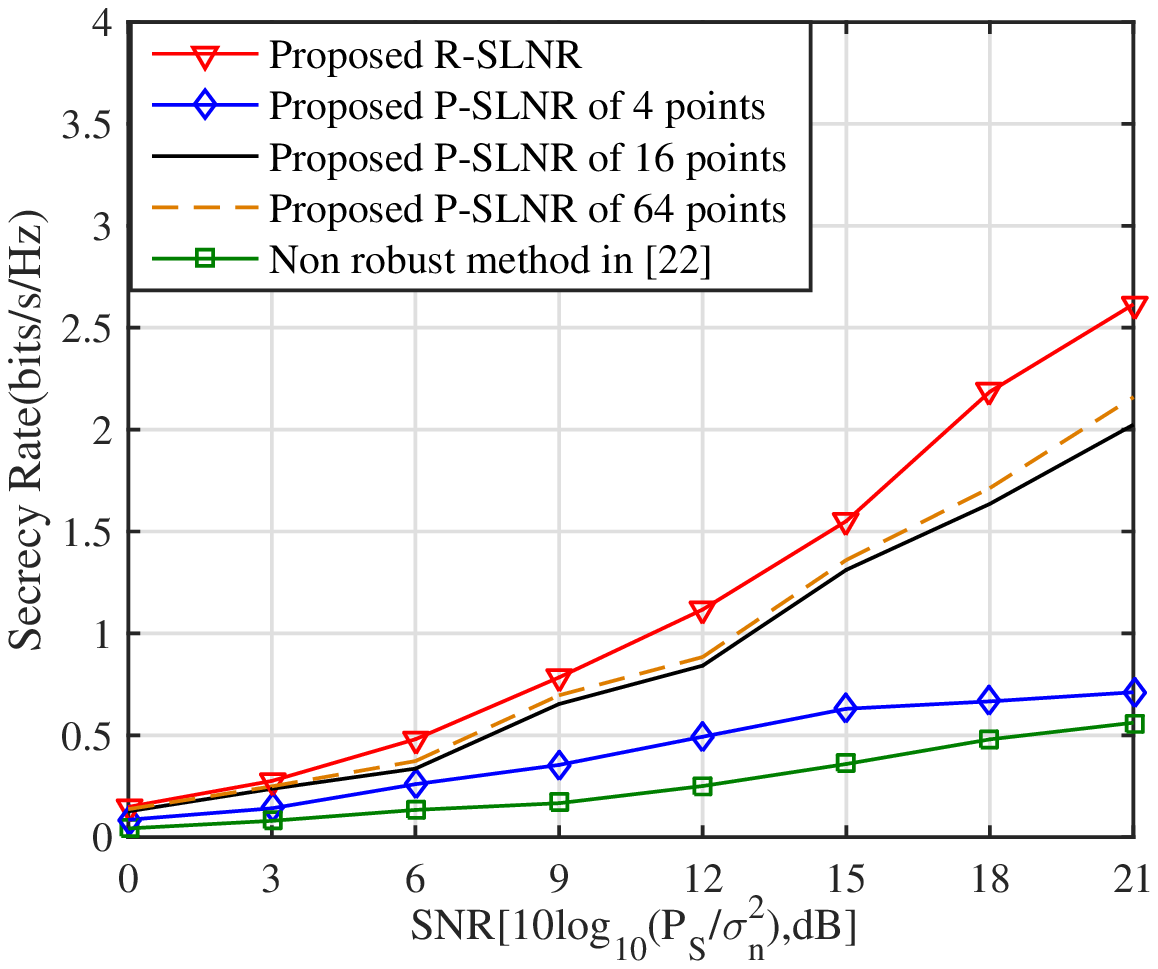}}
\caption{Curves of secrecy rate versus SNR($10log_{10}(P/\sigma_n^2)$) of the proposed methods with $\Delta r_{max}=20m$ and $\Delta r_{max}=100m$.}
\label{SR4}
\end{figure}

We also discuss the secrecy rate performance with different estimation error range. Fig. \ref{SR3} and Fig. \ref{SR4} depict the curve of secrecy rates versus SNR with different maximal estimation direction angle error $\Delta\theta_{max}$ and distance error $\Delta r_{max}$. With the increase of $\Delta\theta_{max}$ and $\Delta r_{max}$, the estimation error region becomes larger, and even larger than the main-lobe which may lead to the same results with that in Fig.\ref{SR1} and Fig. \ref{SR2}. Fig. \ref{angle9.eps} and Fig. \ref{D=100.eps} show that the secrecy rate decreases as the estimation error region becomes larger.

In summary, both the R-SLNR-ANLNR maximization-based scheme and the P-SLNR-ANLNR maximization-based scheme are feasible to achieve robust SPWT, and we can observe that the two schemes are related to the main-lobe size and the estimation error range. Explicitly, when the main-lobe size becomes smaller or the estimation error range becomes larger, the secrecy rate decreases. Besides, compared to the R-SLNR-ANLNR maximization-based scheme, the P-SLNR-ANLNR maximization-based scheme has a comparable secrecy performance but has a lower algorithm complexity, and the curves of the two proposed schemes are almost overlapped, when the main-lobe is large enough or the estimation error is sufficient small.
%
%
%

\section{Conclusion}
In this paper, a R-SLNR maximization-based scheme and a P-SLNR maximization-based scheme are proposed to achieve a regional robust SPWT in a UAV-based multi-user scenario. Based upon these schemes, we obtain the following insights: 1) The proposed schemes can generate a high SINR peak only at the desired regions with the measurement errors, while result in a low flat SINR plane for other eavesdropper regions, which is far less than that at the desired positions formed. 2) The secrecy rate is related to the main lobe size and the estimation error region size, when antenna number $N_T$ and signal bandwidth $B$ increase which lead the main lobe to be smaller, the secrecy rate reduced. Also when maximal error $\Delta\theta_{max}$ and $\Delta r_{max}$ increase which lead the error region to be larger, the secrecy rate reduced too. 3) Since the covariance matrix $\hat{\mathbf{R}}_{{s_m}}$ is a sum for several points instead of an integration for the whole estimation region in the R-SLNR maximization-based scheme, the P-SLNR maximization-based scheme has low complexity, and the secrecy performance is comparable to that of the P-SLNR maximization-based scheme. As the main lobe size shrinks or the estimation error region size enlarges, more points are needed for the P-SLNR maximization-based scheme to guarantee the secrecy performance approaching the regional SLNR maximization-based scheme. Due to the low-complexity and high security properties of proposed scheme, it will be potentially applied in the future scenarios such as unmanned aerial vehicle communications, intelligent connected vehicle communications and so on.

\ifCLASSOPTIONcaptionsoff
  \newpage
\fi
\bibliographystyle{IEEEtran}

\bibliography{IEEEfull,reference}

\end{document}